\documentclass[12pt,fleqn]{article}
\usepackage{epic,eepic}
\usepackage{graphicx}
\usepackage{epstopdf}
\usepackage{pdftricks}
\usepackage{lineno}
\begin{document}
\title{Thermodynamic geometry of the Gaussian core model fluid}

\author{George Ruppeiner\footnote{New College of Florida, Sarasota, Florida, USA}, Peter Mausbach\footnote{Technical University of Cologne, Cologne, Germany (pmausb@gmx.net)} \footnote{Corresponding author}, and Helge-Otmar May\footnote{University of Applied Sciences, Darmstadt, Germany, deceased}}

\date{\today}
\maketitle

\begin{abstract}

The three-dimensional Gaussian core model (GCM) for soft-matter systems has repulsive interparticle interaction potential $\phi (r) = \varepsilon\, {\rm exp}\left[ -(r/\sigma)^{2} \right]$, with $r$ the distance between a pair of atoms, and the positive constants $\varepsilon$ and $\sigma$ setting the energy and length scales, respectively. $\phi (r)$ is mostly soft in character, without the typical hard core present in fluid models. We work out the thermodynamic Ricci curvature scalar $R$ for the GCM, with particular attention to the sign of $R$, which, based on previous results, is expected to be positive/negative for microscopic interactions repulsive/attractive. Over most of the thermodynamic phase space, $R$ is found to be positive, with values of the order of $\sigma^3$. However, for low densities and temperatures, the GCM potential takes on the character of a hard-sphere repulsive system, and $R$ is found to have an anomalous negative sign. Such a sign was also found earlier in inverse power law potentials in the hard-sphere limit, and seems to be a persistent feature of hard-sphere models.

\end{abstract}


\section{Introduction}
The relatively new approach of Riemannian thermodynamic geometry, with the thermodynamic Ricci curvature scalar $R$ of central interest \cite{Rupp95_96}, has attracted some recent attention. $R$ is a geometric invariant in equilibrium thermodynamics containing information about the intrinsic physics at the mesoscopic level. It's connection with the character of the microscopic interactions is perhaps the most significant result of this approach. The sign of $R$ seems to indicate whether the ``effective" interaction is predominantly attractive ($R < 0$) or repulsive ($R > 0$) \cite{Rupp10, Rupp16}, in the curvature sign convention of Weinberg \cite{Weinberg1972}. The magnitude $|R|$ has been interpreted as a measure of the size of mesoscopic structures, which near a critical point is the correlation length.

\par
In this paper, we evaluate $R$ for the three-dimensional (3D) Gaussian core model (GCM), characterized by a repulsive interparticle pair interaction potential $\phi(r)$ decreasing in strength with distance $r$ between atom pairs as a Gaussian function:

\begin{equation}
\phi(r) = \varepsilon\,{\rm exp}\left[ -\left(\frac{r}{\sigma}\right)^{2} \right],
\label{EqGCD}
\end{equation}

\noindent with the positive constants $\varepsilon$ and $\sigma$ setting the energy and length scales, respectively. This model clearly lacks the repulsive hard-cores typically present in fluid state models. It is thus useful for modeling systems consisting of large soft molecules, like polymers, without a clear center \cite{Stillinger1976, Stillinger1997}.

\par
There have been numerous applications of thermodynamic geometry to problems, including Ising based models \cite{Ruppeiner1981, Janyszek1989, Janke02, Janke03, Mirza2013, Ruppeiner2015a}, Bose and Fermi gases \cite{Janys90, Oshima1999, Mirza2009, Ubriaco13}, and quantum chromodynamics \cite{Castor18}. In addition, the Riemannian geometry has become increasingly important in the field of fluid thermodynamics. Evaluated fluid models include van der Waals and Lennard-Jones (LJ) \cite{Ruppeiner12a, MayMaus12a, MayMauRupp13, MauKoeVra18}. In addition, studies were performed for a large number of real fluids \cite{Ruppeiner12b, RuppMauMay15, Rupp17}. Other interesting phenomena have been explored with the geometry of thermodynamics. Supercooled water \cite{MayMauRupp15, Mausbach2019} near a conjectured second critical point in the metastable liquid phase is one such case. More recently, the thermodynamic Ricci curvature scalar $R$ has been evaluated for cases having a full three dimensional Riemannian geometry \cite{Ruppeiner1990, Kaviani1999, Erdem2018, Ruppeiner2020}. Thermodynamic geometry has also seen numerous applications to black hole thermodynamics (for recent reviews, see, e.g., \cite{Sahay2017, Rupp18}).

\par
The resulting picture is broad, consistent, and compelling. For most substances in the fluid phase, attractive interparticle interactions dominate, and most fluid states exhibit negative thermodynamic curvature $R$. Attractive interactions also dominate near the critical point, where $R$ generally diverges to negative infinity. For systems without interparticle interactions, such as the single component ideal gas, $R = 0$, and for solids it is expected that $R$ is positive \cite{MayMauRupp13}.

\par
In a recent study Bra\'nka, Pieprzyk, and Heyes (BPH) \cite{Branka2018} found negative values for $R$ for the hard-sphere limit of inverse power law (IP) fluids. Positive $R$ is expected for repulsive interactions. We discuss this BPH anomaly below. The underlying physical reason for these negative $R$ values is not yet clear. But these results seem persistent, extending even to binary fluids \cite{Ruppeiner2020}.

\par
Colloidal suspensions open up another field for exploring unusual phase diagrams. Polymer coils in solution, star polymers \cite{Likos01, Likos06}, and microgels \cite{Heyes09} are typical examples of soft colloids. Microgels fall between near-hard-sphere colloids and ultra-soft colloidal systems like polymer coils and star polymers. The soft-sphere or inverse power (IP) potential \cite{Branka2018, Rupp05} allows a tuning of purely repelling interactions from very soft to extremely hard. This system has frequently been used to interpret experimental data of microgels \cite{Heyes09, Senff99}.

\par
Ultra-soft colloidal particles are fractal objects of fluctuating shape. An example of such systems consists of polymers in a good solvent, where the highly penetrable coils have finite interactions for all distances. As a consequence, these aggregates can be compressed to attain very large packing fractions. Ultra-soft colloids can be modeled by point particles through a bounded soft repulsive potential like the GCM \cite{Stillinger1976, Likos01}. An extension of the GCM leads to the generalized exponential model (GEM) with a potential described by

\begin{equation}
\phi(r) = \varepsilon\, {\rm exp}\left[ -(r/\sigma)^{n} \right],
\label{EqGEM}
\end{equation}

\noindent where $n$ is a softness exponent. The GCM is recovered when $n= 2$ and for $n > 2$ the GEM displays cluster formation \cite{Mladek06, LikMlad07}.

\par
The present study is the first Riemannian geometry analysis of ultra-soft atoms, and it is limited to $n = 2$ (GCM). The thermodynamic and dynamic behavior of a GCM fluid differs significantly from other simple fluids, like LJ \cite{Prestipino2005, MauMay06, MauSad11}. Below a characteristic temperature $T_m$, increasing the number density $\rho$ at constant temperature $T$ from a dilute fluid state leads to a phase transition to a highly ordered solid state, similar to the behavior of ordinary simple fluids. Upon further compression, however, the particles increasingly overlap, and the flat portion of the interaction potential near $r=0$ is approached. Here, the interaction force gradually turns off, the particles decorrelate, and there is a reentrant phase transition from the solid to a fluid state \cite{MauMay06}.

\par
\begin{figure}[htbp!]
\begin{center}
\includegraphics[width=4.0in]{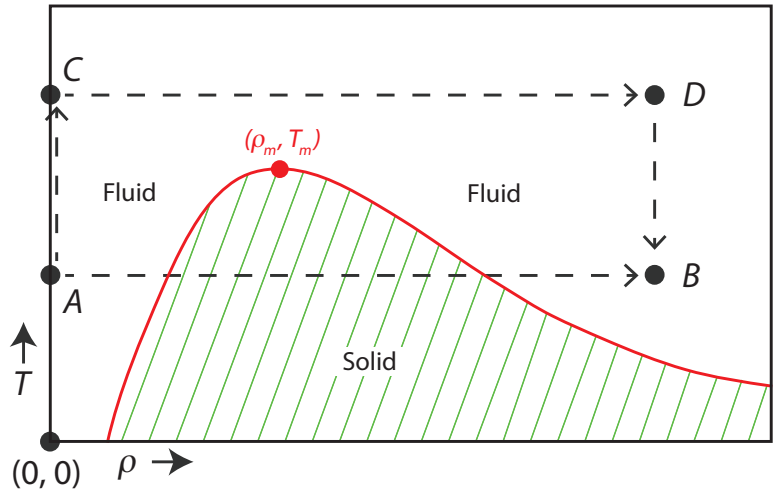}
\end{center}
\caption{A schematic phase diagram for the GCM. The solid red curve indicates the freezing curve for the phase transition from fluid (above the curve) to solid (below the curve). The density discontinuity between the coexisting phases is not shown. The red dot at $(\rho_m,T_m)$ marks the highest temperature on the freezing curve. The dashed arrows represent the two reference curves $AB$ and $ACDB$ connecting the two points $A$ and $B$, discussed in the Appendix.
}
\label{Figure1}
\end{figure}

Figure \ref{Figure1} shows a schematic phase diagram for the GCM, with the point with coordinates $(\rho_m, T_m)$ corresponding to the highest temperature at which a solid phase could exist. The rectangle ABCD will be discussed further below.

\par
Despite the superficial resemblance of Fig. \ref{Figure1} to the phase diagrams of typical fluids (e.g., van der Waals), differences prevail. For example, the highest point $(\rho_m, T_m)$ corresponds to a first-order phase transition, and not to a second-order critical point as in fluids. Proof was presented by Prestipino, et al. \cite{Prestipino2005}, who found a discontinuity $\Delta s\neq 0$ in the molar entropies $s$ between the fluid and solid phases at $(\rho_m, T_m)$. Even though Stillinger \cite{Stillinger1976} proved that the molar volumes $v$ between the fluid and solid phases are equal at $(\rho_m, T_m)$, $\Delta v=0$, a second-order phase transition requires that both $\Delta s$ and $\Delta v$ be zero.

\par
This view is amplified by the Clausius-Clapeyron equation \cite{Landau1980} for the slope of the coexistence curve:

\begin{equation}\frac{dp}{dT}=\frac{\Delta s}{\Delta v}.\label{664}\end{equation}

\noindent As we increase $\rho$ along the coexistence curve, at the highest point $(\rho_m, T_m)$ we have $dT=0$. But with increasing $\rho$, the pressure $p$ also increases, and $dp\neq 0$ at $(\rho_m, T_m)$. The slope in Eq. (\ref{664}) thus goes to infinity at $(\rho_m, T_m)$, consistent with $\Delta s\neq 0$ and $\Delta v=0$. Of course, since $(\rho_m, T_m)$ is not a critical point, there will be no Widom line.

\par
In addition to thermodynamic anomalies, such as a negative thermal expansion coefficient, the GCM displays anomalous transport properties \cite{MayMau09, MauMay09, AhmMauSad09, AhmMauSad10}, leading to the violation of the Stokes-Einstein relation \cite{MayMau07, MauMay08}.

\par
This paper is organized as follows: section \ref{geoanalysis} compiles the necessary formulas for the calculation of the thermodynamic Ricci curvature scalar $R$. In section \ref{results}, a detailed geometric analysis of the considered phase region is presented. This discussion includes an explicit discussion of hard-sphere fluids, which have negative $R$. Conclusions in section \ref{concl} sum up the findings, and an appendix adds some details for determining the thermodynamics of the GCM.

\section{Thermodynamic geometry and analysis \label{geoanalysis}}

For a pure fluid, the internal energy $U$, the number of atoms $N$, the volume $V$, and the entropy $S$ define the fundamental thermodynamic equation $U = U(S, N, V)$ \cite{Callen}. The temperature $T = U,_{S}$, the pressure $p = - U,_{V}$, and the chemical potential $\mu = U,_{N}$, may be obtained directly from the internal energy $U$. Here, the comma notation indicates partial differentiation. The Helmholtz free energy is defined as $A =U - TS$, and the Helmholtz free energy per volume is $f = A/V$. The number density is $\rho = N/V$, and $f$ can be written as $f = f(T, \rho)$, with the entropy per volume $s = S/V = - f,_{T}$, the chemical potential $\mu = f,_{\rho}$, and the internal energy per volume $u =U/V = f + Ts$. The pressure $p = -f + \rho \,f,_{\rho} = -u + Ts + \mu \rho$.

\par
Using $(T, \rho)$ coordinates as state variables, a thermodynamic line element $d \ell^{2}$ can be obtained from the thermodynamic entropy information metric \cite{Rupp95_96}

\begin{equation}
d \ell^{2} = \frac{1}{k_{\rm B}T} \left (\frac{\partial\,s}{\partial\,T} \right )_{\rho} d\,T^{2} + \frac{1}{k_{\rm B}T} \left (\frac{\partial\,\mu}{\partial\,\rho} \right)_{T} d \rho^{2},
\label{metric}
\end{equation}

\noindent where $k_{\rm B}$ is Boltzmann's constant. The thermodynamic metric Eq. (\ref{metric}) induces the thermodynamic Ricci curvature scalar

\begin{equation}
R = \frac{1}{\sqrt g}\,\left[ \frac{\partial}{\partial\,T} \left(\frac{1}{\sqrt g}\,\frac{\partial\,g_{\rho\rho}}{\partial\,T}\right)+ \frac{\partial}{\partial\,\rho} \left(\frac{1}{\sqrt g}\,\frac{\partial\,g_{TT}}{\partial\,\rho}\right)\right],
\label{RiemCurv}
\end{equation}

\noindent with metric elements

\begin{equation}
g_{TT} = -\frac{1}{k_B T}\,\left(\frac{\partial^{\,2}f}{\partial\,T^{2}}\right),
\label{MetrT}
\end{equation}

\noindent and

\begin{equation}
g_{\rho\rho} = \frac{1}{k_B T}\,\left(\frac{\partial^{\,2}f}{\partial\,\rho^{2}}\right),
\label{MetrRho}
\end{equation}

\noindent and with metric determinant

\begin{equation}
g = g_{TT}\;g_{\rho\rho}.
\label{MetrDet}
\end{equation}

\par
To evaluate the thermodynamic geometry in $(T,\rho)$ coordinates requires that we know the fundamental equation $f(T,\rho)$. May and Mausbach \cite{MayMau12b} found $f(T,\rho)$ for the GCM by fitting the extensive computer simulation data by Mausbach and Sadus \cite{MauSad11}. These computer simulations consisted of determining triplet values $(T,\rho,p)$ on a grid. Here, the equation of state (EOS) fit expresses $p=p(T,\rho)$.

\par
May and Mausbach \cite{MayMau12b} worked in reduced units with dimensionless quantities $T'=k_B T/\varepsilon$ and $\rho'=\sigma^3\rho$. With these dimensionless units, all the other thermodynamic properties can be made consistently and uniquely dimensionless using $\varepsilon$ and $\sigma$. However, although we feature the GCM in this paper, we also have other models in play, each with it own options for working in dimensionless units. To avoid confusion about what units are employed, we will simply work mostly with real units, with our only concession being to set $k_B=1$. The computer simulations covered number densities ranging from $\sigma^3\rho = 0.004$ to $2.0$, and temperatures ranging from $k_B T/\varepsilon = 0.002$ to $3.0$. However, the EOS was only fitted up to $k_B T/\varepsilon = 0.08$. The solid-state region of the GCM model was omitted from the fit.

\par
Pressure and internal energy data obtained from the simulations \cite{MauSad11} were used to correlate an empirical ansatz function \cite{MayMau12b} for the pressure

\begin{equation} p(T, \rho) = p^{id}(T, \rho) + p^{ex}(T, \rho).\label{press}\end{equation}

\noindent Here $p^{id}(T, \rho) = \rho\,T$ is the purely ideal gas part of the pressure, and $p^{ex}(T, \rho)$ is the excess part contributed by the microscopic interactions. The internal energy per volume is related to the pressure by

\begin{equation}
u(T,\rho) = \frac{3}{2}\,\rho\,T + \rho\, \int^\rho_{\rho=0} \left[ p^{ex} - T\left( \frac{\partial\,p^{ex}}{\partial\,T}\right) \right] \frac{d\rho}{\rho^{2}}.
\label{intenerg}
\end{equation}

\noindent The entropy per volume is

\begin{equation}
s(T,\rho) =c \rho + \frac{3}{2}\,\rho\,{\rm ln}\, T - \rho\,{\rm ln}\,\rho - \rho\,\int^\rho_{\rho=0} \left(\frac{\partial\,p^{ex}}{\partial\,T} \right) \frac{d \rho}{\rho^{2}}.
\label{intenerg}
\end{equation}

\noindent Here, $c$ is a constant that does not ultimately appear in the expression for $R$. From these expressions, the Helmholtz free energy per volume $f(T,\rho) = u - T s$ can be calculated. The appendix adds some detail to this formalism.

\par
Most of our investigations are applicable up to moderately high densities and temperatures, where we have good computer simulation data for $p^{ex}(T, \rho)$, as well as a good coexistence curve. But also amenable to analysis is the very high density and temperature regime, where the finite core of the potential leads to an overlap of atoms, and ideal-gas like behavior is expected \cite{Lang00, Louis00}.

\par
In the very high density and temperature regime, the thermodynamics of the GCM fluid is described remarkably well by the random phase approximation (RPA) \cite{Lang00, Louis00}, where the Helmholtz free energy per volume becomes \cite{Lang00, Louis00, Ikeda11, Frydel16}

\begin{equation}
f(T, \rho) = f^{id}(T, \rho) + \frac{\pi^{3/2} \rho^{2}\, \sigma^{3}\, \varepsilon}{2} - \frac{\rho\, \varepsilon}{2} \left[1 + \frac{1}{\gamma} {\rm Li}_{5/2}(-\gamma)\right].
\label{HFEnergyRPA}
\end{equation}

\noindent The first term in Eq. (\ref{HFEnergyRPA}) is the ideal-gas contribution
\begin{equation}
f^{id}(T, \rho) = \rho \,T \left\{ \rm{ln} \left[ \rho\, \Lambda^{3} \right ] - 1 \right\},
\label{HFEidgas}
\end{equation}

\noindent the second term is the uniform part, and the third term is the RPA fluctuation part. $\Lambda = \Lambda(T)$ is the thermal de Broglie wave length,

\begin{equation}
\Lambda=\frac{h}{\sqrt{2\pi m T}},
\end{equation}

\begin{equation}
\gamma = \frac{ \pi^{3/2} \rho\, \sigma^{3}\, \varepsilon}{T} = \pi^{3/2}\,\frac{\rho'}{T'}
\label{couplPar}
\end{equation}

\noindent is a dimensionless coupling parameter, and

\begin{equation}
{\rm Li}_n(x) = \sum_{k=1}^\infty \frac{x^{k}}{k^{n}} \;
\label{Polylog}
\end{equation}

\noindent is the $n$'th polylogarithm. From Eqs. (\ref{RiemCurv}) and (\ref{HFEnergyRPA}) we may obtain the thermodynamic Ricci curvature scalar $R$.

\par
We may also get other quantities of interest. For example, the excess pressure is
\begin{equation}
p^{ex}(T, \rho) = \frac{\pi^{3/2} \rho^{2}\, \sigma^{3}\, \varepsilon}{2} - \frac{\rho\, \varepsilon}{2 \gamma} \left[\,{\rm Li}_{3/2}(-\gamma) - {\rm Li}_{5/2}(-\gamma)\,\right ],
\label{pressExRPA}
\end{equation}

\noindent and the excess chemical potential is

\begin{equation}
\mu^{ex}(T, \rho) = \pi^{3/2} \rho \, \varepsilon\, \sigma^{3} - \frac{\varepsilon}{2} \left[1 + \frac{1}{\gamma} {\rm Li}_{3/2}(-\gamma)\right].
\label{ChePotEx}
\end{equation}

\par
In addition, the radial distribution function at a particle distance $r=0$ from some atom can be calculated analytically within RPA as

\begin{equation}
g_{\,\rm RPA}(r = 0) = 1 + \frac{1}{\pi^{3/2} \rho\, \sigma^{3}}\, {\rm Li}_{3/2}(-\gamma)\, .
\label{gRPA}
\end{equation}

\noindent This expression may be viewed as a measure of the extent to which the GCM particles overlap. The disappearance of the correlation hole in $g_{\,\rm RPA}(r)$, characterized by an approach $g_{\,\rm RPA}(r = 0) \to 1$ as $T \to \infty$ and $\rho \to \infty$, signals the occurrence of ideal gas-like behavior. At low temperatures, Eq. (\ref{gRPA}) can lead to an unphysical negative $g_{\,\rm RPA}(r = 0)$. The corresponding threshold temperature $T_{\rm th}(\rho)$ is frequently discussed as a limit that restricts the validity of the RPA \cite{Ikeda11}. In this study $T_{\rm th}(\rho)$ is determined by solving Eq. (\ref{gRPA}) numerically, on setting $g_{\,\rm RPA}(r = 0; \rho, T = T_{\rm th}) = 0$. The course of $T_{\rm th}(\rho)$ is displayed in Figure \ref{RPAcontour} below.

\par
\begin{figure}
\begin{minipage}[b]{0.5\linewidth}
\includegraphics[width=2.7in]{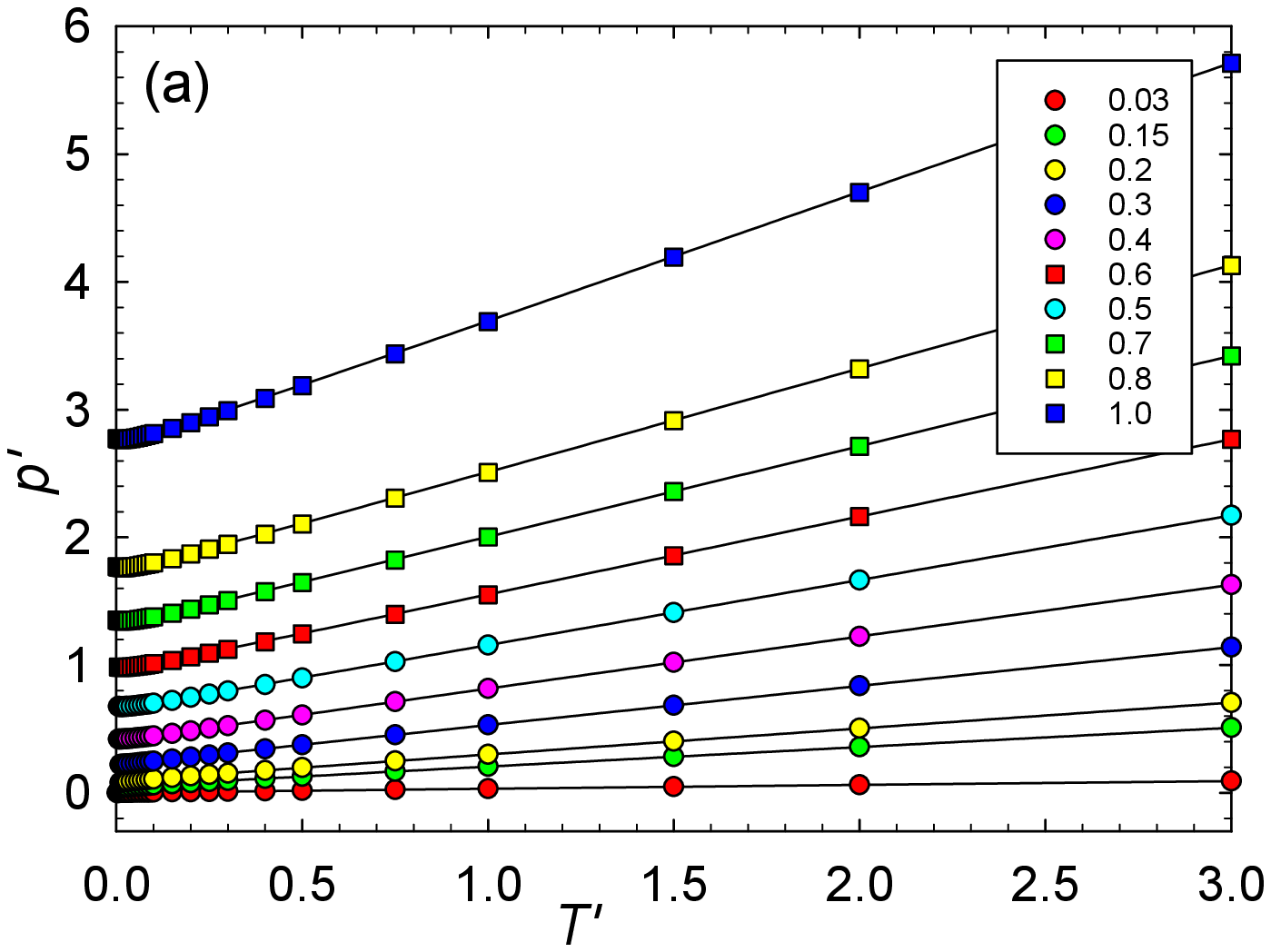}
\end{minipage}
\hspace{0.0 cm}
\begin{minipage}[b]{0.5\linewidth}
\includegraphics[width=2.75in]{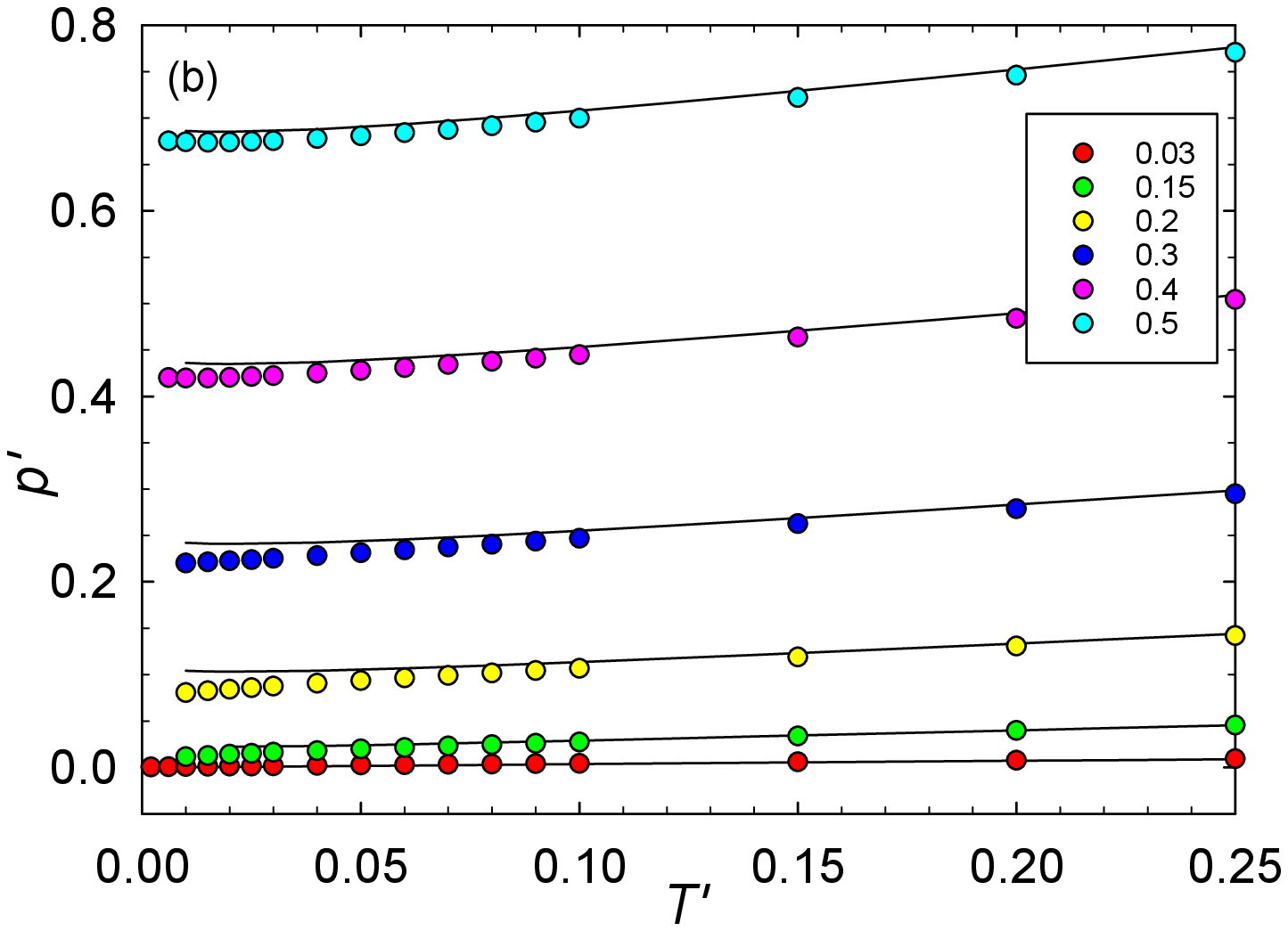}
\end{minipage}
\caption{Comparison of the temperature-dependent pressure isochores between simulation results from Mausbach and Sadus \cite{MauSad11} (symbols) and the RPA according to the Eqs. (\ref{press}) and (\ref{pressExRPA}) (solid lines).
(a) shows pressure isochores ranging from $\rho' = 0.03$ to $1.0$, capturing relatively high temperatures, and (b) focusses on low temperatures for densities from $\rho' = 0.03$ to $0.5$.
}
\label{Pressure}
\end{figure}

In order to assess the quality of the RPA, pressure isochores are compared in Figure \ref{Pressure} between simulation results obtained from Mausbach and Sadus \cite{MauSad11} and those of the RPA obtained from Eq. (\ref{pressExRPA}). The agreement between RPA and simulation results is very good for high temperatures as shown in Fig. \ref{Pressure}(a). For low temperatures, however, there are deviations from the simulation results, as shown in Fig. \ref{Pressure}(b). Note that the threshold temperatures $T_{\rm th}(\rho)$ lie outside the temperature axis of Fig. \ref{Pressure}(b), suggesting that the RPA pressure apparently also shows good results for $T < T_{\rm th}$.

\section{Results and discussion \label{results}}

In this section, we present our results. For the following discussion, we label the region with $\rho'<\rho_m'$ (=0.24265) as the low density side (LDS) of the phase maximum, and the region with $\rho'>\rho_m'$ as the high density side (HDS) \cite{MauAhmSad09}.

\subsection{General overview}

\begin{figure}
\begin{minipage}[b]{0.5\linewidth}
\includegraphics[width=2.7in]{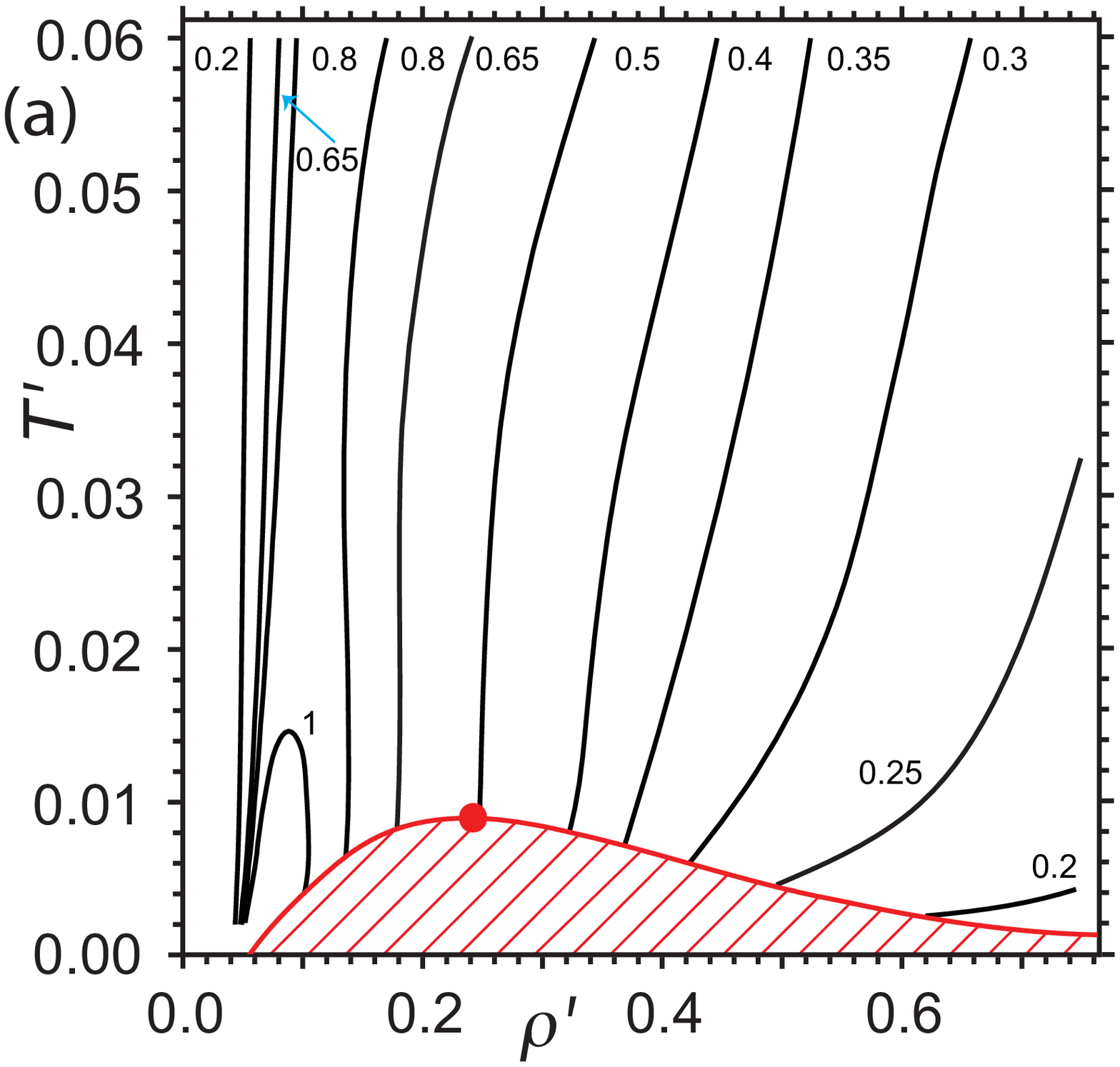}
\end{minipage}
\hspace{0.0 cm}
\begin{minipage}[b]{0.5\linewidth}
\includegraphics[width=2.7in]{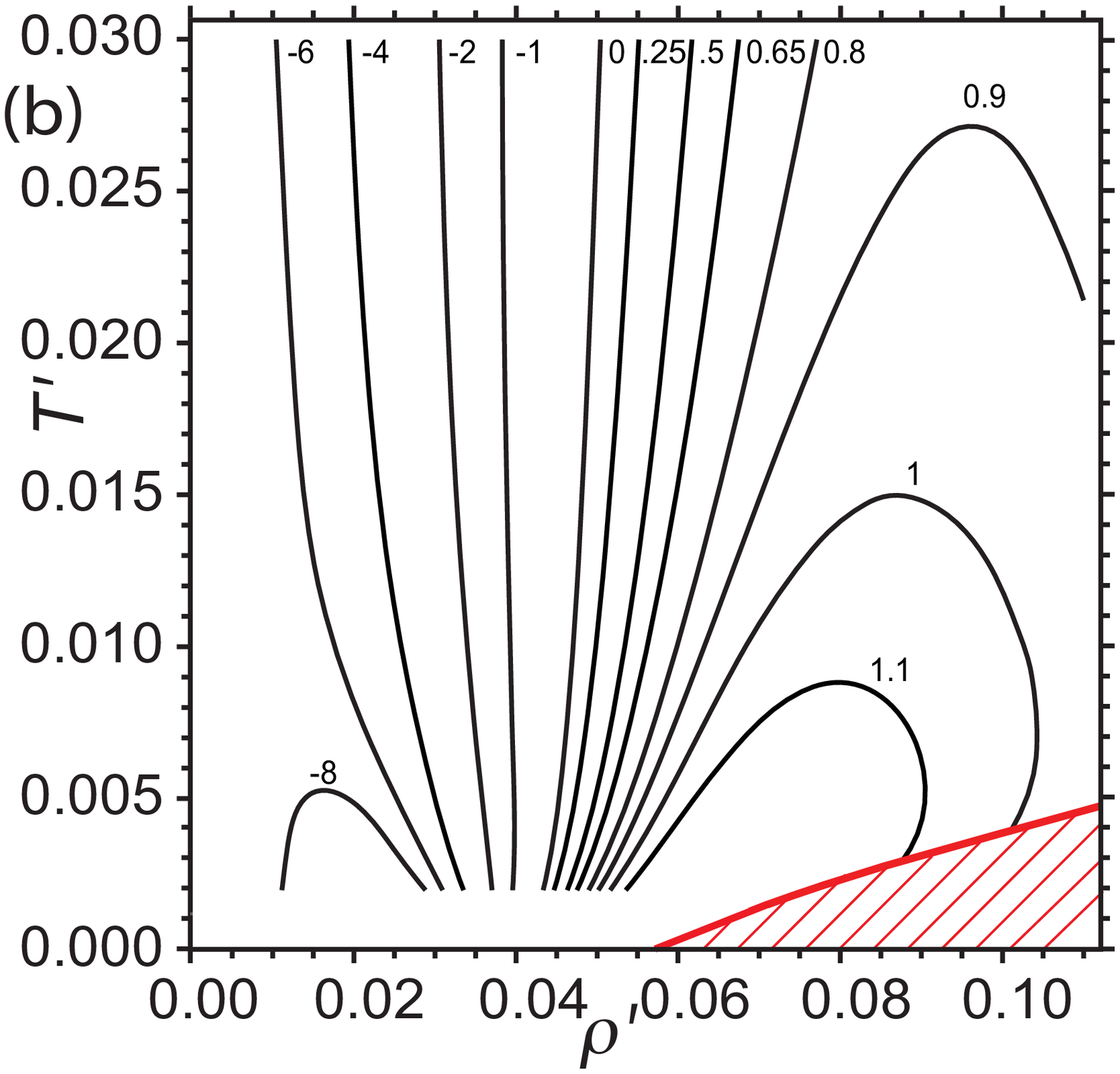}
\end{minipage}
\caption{$R$-diagrams showing the contours (solid black curves) of $R' = R/\sigma^{3}$ in the dimensionless $(\rho',T')$ space. The red curve shows the freezing line of the fluid-solid phase transition, peaking at $(0.24265, 0.00904)$. (a) shows a wide view, capturing most of the simulation data, and (b) focusses on the LDS. Generally, $R$ is mostly positive, and with magnitude less than the molar volume, $R\rho<1$. However, in (b) we see negative values of $R$ getting larger as $\rho$ gets smaller, but with $|R|\rho<1$.
}
\label{Figure3}
\end{figure}

Figure \ref{Figure3} shows the contour lines for the thermodynamic Ricci curvature scalar $R' = R/\sigma^{3}$ over two rectangles in $(\rho',T')$ space in the domain of the computer simulations by May and Mausbach \cite{MayMau12b}. To calculate $R$, we used the fit formula in \cite{MayMau12b} for $p^{ex}$, Eq. (\ref{RiemCurv}) for $R$, and the methods in the Appendix. The coexistence curve was taken from Mausbach et al. \cite{MauAhmSad09}, who computed both the freezing and the melting lines. Fig. \ref{Figure3} shows only the freezing line.

\par
Because the GCM has purely repulsive microscopic interactions, positive $R$ for the whole phase space is expected. Particularly, for large $\rho$, where we expect a condensed fluid state, general expectations are that $R$ tends to be positive, with the dimensionless $|R|\rho\sim +1$ (see Figure 10.4(c) in \cite{Ruppeiner2014}). For GCM, indeed, Fig. \ref{Figure3}(a) shows that we mostly have $R$ roughly in this zone. The exception is the region at very low density $\rho' < 0.04$ in Fig. \ref{Figure3}(b), where there are significant instances of negative $R$.

\par
The low density and low temperature GCM regime was identified by Stillinger \cite{Stillinger1976} as having effectively very ``hard'' repulsive interactions between atoms. By ``hard'' we mean an interaction potential of zero/infinity for a pair of atoms separated by a distance bigger/less than some interaction range. We thus naturally associate the GCM's negative $R$ regime with the BPH anomaly \cite{Branka2018}, as we discuss in subsection \ref{hardcore}.

\par
The behavior of $R$ close to the point at $(\rho_m, T_m)$, marked as a red dot in Fig. \ref{Figure3}(a), is of special interest. At a critical point, $R$ generally diverges to negative infinity. Since $R$ remains finite near $(\rho_m, T_m)$, our geometry analysis of the GCM fluid also suggests that this point is not a solid-fluid critical point.

\subsection{High-temperature limit (HTL)}

\par
The fitted EOS data are restricted to $T'<0.08$, so at this point we can only speculate on the behavior of $R$ for higher temperatures. Naively, we would expect that as $T$ increases at constant $\rho$, the atoms move faster, and all effects of the microscopic interactions decrease. In this event, we expect an approach to ideal gas behavior, with $R=0$ \cite{Ruppeiner1979}. The behavior of the RPA radial distribution function $g_{\,\rm RPA}(r = 0)$ shown in Eq. (\ref{gRPA}) supports this ideal gas like behavior at high temperatures.

\par
\begin{figure}[htbp!]
\begin{center}
\includegraphics[width=4.0in]{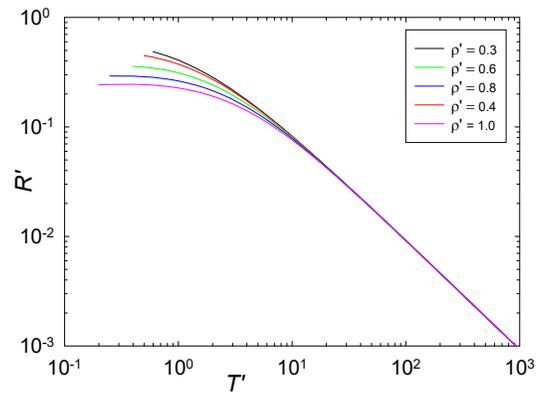}
\end{center}
\caption{\label{hightemp} Temperature-dependent thermodynamic Ricci curvature scalar $R' = R/\sigma^{3}$ along isochores between $\rho' = 0.3$ and $1.0$ according to the RPA, Eq. (\ref{HFEnergyRPA}). For large $T'$, $R'$ and $T'$ are related as a power law, independent of $\rho'$.
} 
\end{figure}

Fig. \ref{hightemp} shows $R' = R/\sigma^{3}$ along five isochores between $\rho' = 0.3$ and $1.0$ obtained from the formulas of the RPA. Clearly, $R'$ approaches zero at high temperatures $T'$ as a power law.. The curves start roughly at $T'_{\rm th}$, the temperature with a positive $g_{\,\rm RPA}(r = 0)$ \cite{Ikeda11}. However, the gap between $T' = 0.08$ and $T'_{\rm th}$ remains unexplored by either the fitted EOS data or the RPA.

\par
A possible scenario for the $R$ contours represented in Fig. \ref{Figure3}(a) is then that each contour rises with increasing $T$, reaches a peak, and then comes back down. The smaller the positive values of $R$, the further up the contours go. In Fig. \ref{Figure3}(b), such behavior is explicitly shown by the contours with $R'= $ $0.9$, $1.0$, and $1.1$. If this indeed the pattern, other contours presumably peak at a $T$ higher than the values probed by the fitted EOS data, probably within the gap $0.08 < T' < T'_{\rm th}$. The behavior for the positive $R$ contours displayed are consistent with this possibility that high $T$ corresponds to ideal gas like behavior $R\to 0$.

\subsection{High-density limit (HDL)}

\par
The high-density limit of the GCM fluid state requires special consideration. Here, the bounded character of the GCM potential results in weaker structural correlations. The system approaches a kind of ``infinite density ideal gas" limit, again with pair-correlation function $g(r) = 1$. Stillinger \cite{Stillinger1976} assumed that the internal energy of the high-density GCM fluid can be approximated by that of a random distribution of points interacting by means of the GCM potential, where the positions of the points are uncorrelated. Investigations of this behavior by means of a geometric analysis is of some interest.

\par
The fact that the ideal gas has a curvature of $R = 0$ gave rise to the interpretation that $R$ is a measure of the interaction strength. Due to vanishing interactions at high densities, the result $R = 0$ should also be found for the GCM fluid in the HDL.

\par
\begin{figure}[htbp!]
\begin{center}
\includegraphics[width=4.0in]{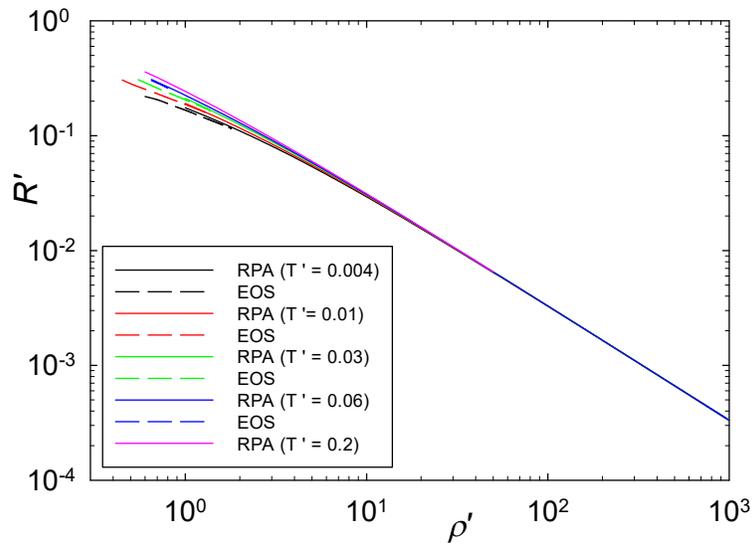}
\end{center}

\caption{\label{highdens} Density dependent thermodynamic Ricci curvature scalar $R' = R/\sigma^{3}$ along isotherms between $T' = 0.004$ and $0.2$ according to the RPA, Eq. (\ref{HFEnergyRPA}) (solid lines), and between $T' = 0.004$ and $0.06$ from the EOS, ref. \cite{MayMau12b} (dashed lines). For large $\rho'$, $R'$ and $\rho'$ are related as a power law, independent of $T'$.}
\end{figure}

Figure \ref{highdens} shows $R'$ along five isotherms between $T' = 0.004$ and $0.2$ obtained from the formulas of the RPA. Furthermore, four isotherms obtained from the EOS \cite{MayMau12b} are shown as dashed lines between $T' = 0.004$ and $0.06$. The curves from both exhibit a smooth connection without significant discontinuities at the point of switching. This may be considered as a good validation of the calculated $R$'s. The curvature becomes $R = 0$ at the infinite density limit and thus, similar to the HTL, confirms the picture developed in thermodynamic geometry. The limiting behavior appears to be a power law, independent of $T'$.

\par
\begin{figure}[htbp!]
\begin{center}
\includegraphics[width=4.0in]{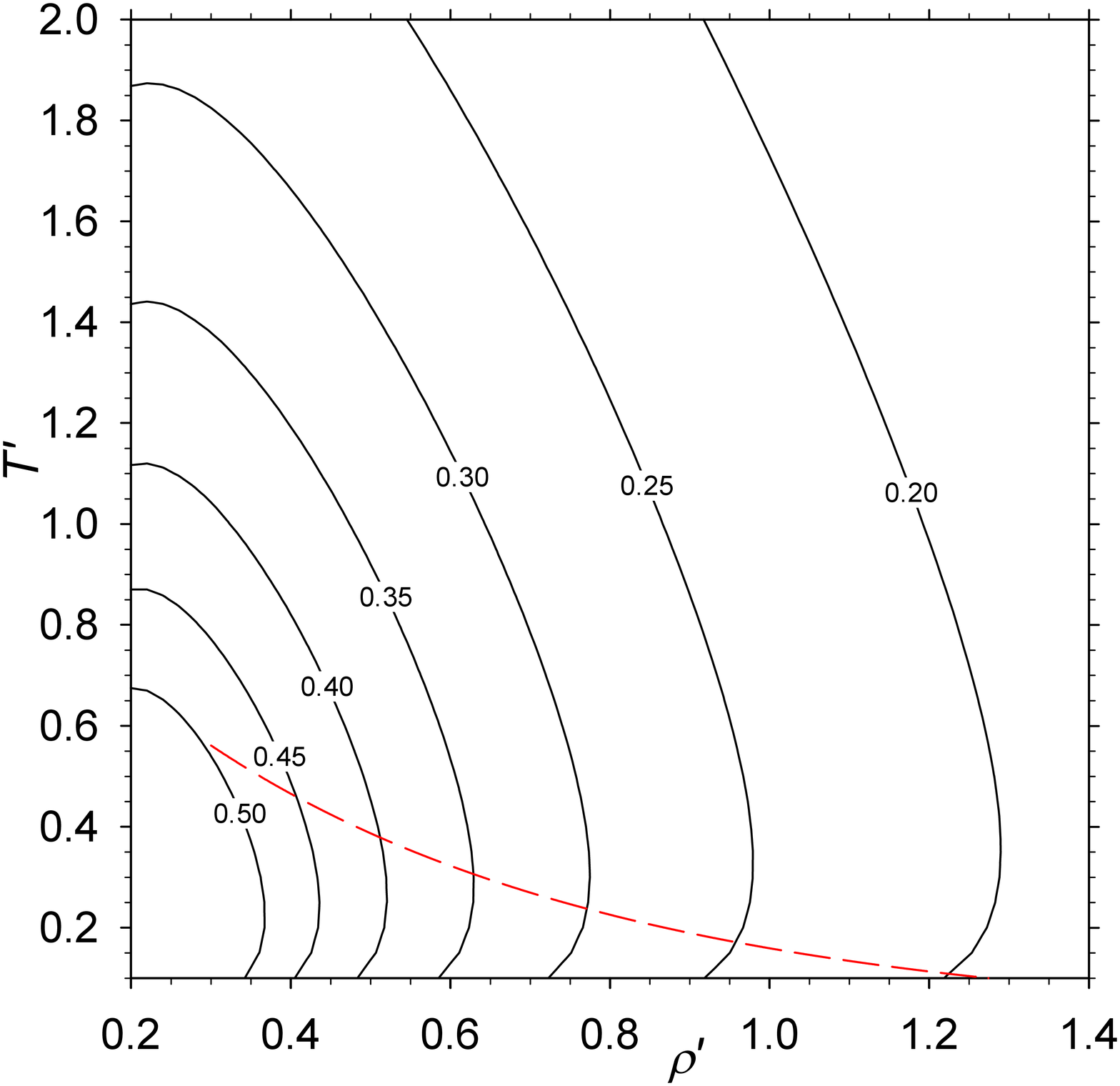}
\end{center}
\caption{\label{RPAcontour} $R$-diagram showing the contours (solid black curves) of $R' = R/\sigma^{3}$ in the dimensionless $(\rho',T')$ space according to the RPA. The red dashed curve shows the course of the $T'_{\rm th}(\rho')$ line. Above this curve, $g_{\,\rm RPA}(r = 0) > 0$, and below it is the unphysical $g_{\,\rm RPA}(r = 0)< 0$.} 
\end{figure}

The overall behavior of the RPA curvature for the GCM fluid is depicted in Figure \ref{RPAcontour}, showing contour lines of $R' = R/\sigma^{3}$ in the $(\rho',T')$ projection. The transition from the HT to the HD regime is continuous. The red dashed curve shows the course of the threshold temperature $T_{\rm th}(\rho)$, such that above this curve $g_{\,\rm RPA}(r = 0) > 0$, and below this curve $g_{\,\rm RPA}(r = 0)< 0$. The curve rises with decreasing density giving a rough indication down to which temperature the RPA can be used.

\par
\subsection{Low density limit (LDL): hard-core regime \label{hardcore}}

At low temperatures ($T\ll\varepsilon$), and with density well under the hard-sphere close packing limit, Stillinger \cite{Stillinger1976, Stillinger1997} showed that the GCM acts like a gas of hard-spheres, with diameter

\begin{equation}\sigma_{gcm}(T)=\sigma\left[\ln\left(\frac{\varepsilon}{T}\right)\right]^{1/2}.\label{24681}\end{equation}

\noindent To see this, consider the pair Boltzmann factor $B(r,T)$ contributed by each pair of interacting particles separated by a distance $r$:

\begin{equation} B(r,T)=\exp\left\{-\frac{\varepsilon}{T}\,\exp\left[-\left(\frac{r}{\sigma}\right)^2\right]\right\}.\end{equation}

\begin{figure}[htbp!]
\begin{center}
\includegraphics[width=4.0in]{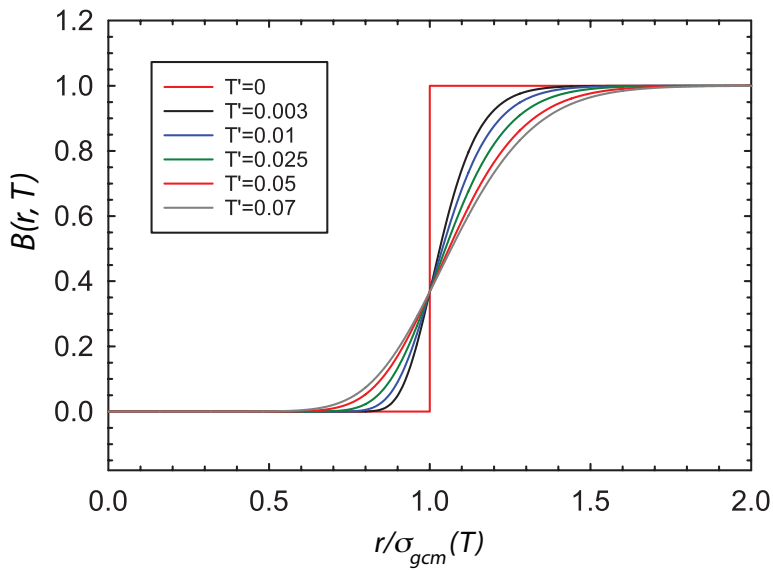}
\end{center}
\caption{The Boltzmann factor $B(r,T)$ versus $r/\sigma_{gcm}(T)$, for five values of $T'=T/\varepsilon$. Also shown is the curve for $T'=0$. }
\label{Figure7}
\end{figure}

\noindent As $T$ decreases towards $0$, $B(r,T)$ goes to $0$ if $r<\sigma_{gcm}(T)$, and $1$ if $r>\sigma_{gcm}(T)$, consistent with $\sigma_{gcm}(T)$ in the role of a hard-sphere diameter. Figure \ref{Figure7} shows $B(r,T)$ for five values of $T'=T/\varepsilon$ in the range of the simulation data. Clearly, none of these values for $T/\varepsilon$ is low enough to result in a well-defined hard-sphere gas.

\par
For context, consider the Carnahan-Starling (CS) equation of state \cite{Carnahan1969}, which is an excellent approximation for the hard-sphere gas. For CS, the atoms have a diameter $\sigma_{cs}$, and the interaction potential for a pair of spheres with centers separated by a distance $r$ is

\begin{equation}\phi(r)=\left\{\begin{array}{ll} \infty \,\, \mbox{if} \,\, r<\sigma_{cs}\\\\ 0\,\,\,\,\,\mbox{if}\,\, r>\sigma_{cs}. \end{array}\right. \label{65843}\end{equation}

\noindent The pressure in the fluid phase is given by the CS equation

\begin{equation} \frac{p V}{N T}=\frac{1+\eta+\eta ^2-\eta ^3}{(1-\eta )^3},\label{CSPressure}\end{equation}
 
\noindent where the dimensionless variable

\begin{equation} \eta=\frac{1}{6}\pi\,\sigma_{cs}^3\,\rho. \end{equation}

\noindent From Eq. (\ref{CSPressure}), we have $\eta\in[0,1)$. The CS equation is in excellent agreement with the virial series expression by Clisby and McCoy \cite{McCoy(2006)}.

\par
\begin{figure}[htbp!]
\begin{center}
\includegraphics[width=4.0in]{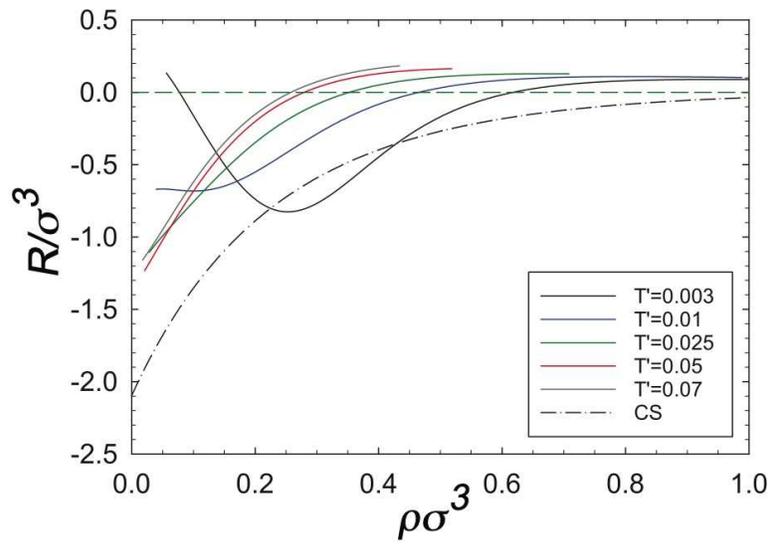}
\end{center}
\caption{$R/\sigma^3$ as functions of $\rho\,\sigma^3$ for five isotherms according to the GCM-EOS \cite{MayMau12b}, with $\sigma=\sigma_{gcm}(T)$. Also shown is the temperature-independent curve of the CS-curvature from Eq. (\ref{CS_EOS}), with $\sigma=\sigma_{cs}$. As could have been anticipated from Figure \ref{Figure7}, agreement between GSM and CS is at most qualitative in this regime.}
\label{Figure8}
\end{figure}

Eqs. (\ref{RiemCurv}) and (\ref{CSPressure}) yield

\begin{equation} \frac{R}{\sigma_{cs}^3}=-\frac{\pi(1-\eta)^3 \left(2 + 5 \eta -\eta ^2\right)}{3\left(1 + 4\eta + 4\eta ^2 - 4\eta ^3 + \eta^4 \right)^2},\label{CS_EOS}\end{equation}

\noindent an expression independent of $T$ \cite{Arenas2020}. $R$ is negative in the range $\eta\in[0,1)$, demonstrating the BPH anomaly. Figure \ref {Figure8} shows $R/\sigma_{cs}^3$ as a function of $\rho\,\sigma_{cs}^3$.

\par
Bra\'nka et al. \cite{Branka2018} employed analytical means of evaluating $R$, with the inverse repulsive power law (IP) pair interaction potential

\begin{equation}\phi(r)=\varepsilon_{ip}\left(\frac{\sigma_{ip}}{r}\right)^{n},\label{746591}\end{equation}

\noindent where $\varepsilon_{ip}$, $\sigma_{ip}$, and $n$ are positive constants. With $n>3$, these repulsive IP systems have a thermodynamic limit, including thermodynamic stability \cite{Ruelle1999}. For soft IP potentials, with $n$ not too big, Bra\'nka et al. \cite{Branka2018} found positive $R$, as expected for repulsive interactions. As $n\to\infty$, the IP potential becomes the hard-sphere potential in Eq. (\ref{65843}), with diameter $\sigma_{ip}$. Here $R$ has the anomalous negative sign, as shown in Eq. (\ref{CS_EOS}). Earlier, Ruppeiner \cite{Rupp05} also reported negative $R$ for instances of the IP systems, but without using the more or less exact means available to Bra\'nka et al. \cite{Branka2018}, and prior to any clear general statement linking the sign of $R$ to microscopic interactions.

\par
In Figure \ref{Figure8} we also show $R$ versus $\rho$ for the GCM for five different values of $T/\varepsilon\ll 1$. The scaling employed is identical in style to that for CS, but with the scaling factor $\sigma_{gcm}(T)$ for the GCM. In this plotting regime, $R$ is also mostly negative for the GCM, but the agreement with the scaled CS $R$ is only qualitative. With the deviation of the pair Boltzmann factors shown in Figure \ref{Figure7} from the hard-core values, we could perhaps do no better. Improvement calls for simulations with smaller values of $T$.

\subsection{Further discussion of the BPH anomaly}

The sign of $R$ is positive over most of the thermodynamic regime of the GCM. $|R|^{1/3}$ in this positive $R$ regime is on the order of the interaction range $\sigma$. These characteristics of solid-like values are found elsewhere: for the Weeks-Chandler-Anderson ansatz \cite{MayMauRupp13}, and in fluids at high temperatures and pressures near the melting line \cite{RuppMauMay15}. It was also found in IP systems for soft potentials \cite{Branka2018}. In addition, we have the canonical example of a repulsive system with positive $R$ in the form of the ideal Fermi gas \cite{Janys90}, where interactions are effectively repulsive by virtue of quantum statistics.

\par
In the low density, low temperature regime of the GCM, we found an anomalous negative $R$. We argued that this is consistent with what has been found in systems with hard-repulsive interactions such as the hard IP potential \cite{Branka2018}. Perhaps the canonical example of such anomalous behavior is the gas of hard spheres. Another example with instances of negative $R$ and repulsive interactions is the binary van der Waals fluid, which has a 3D thermodynamic phase space \cite{Ruppeiner2020}.

\par
How do we interpret this BPH anomaly physically? First, let us say that the interpretation of the sign of $R$ has us in a theoretical ``no man's land,'' with little yet in terms of a mesoscopic theory to justify it. The interpretation is based on model calculations, and on fits to real and computer fluid data. The only relatively firm mesoscopic theoretical result for the physical interpretation of $R$ is $|R|\sim\xi^d$ in critical point regimes. Here, $\xi$ is the correlation length and $d$ is the dimension of the system. This result is based on a covariant thermodynamic fluctuation theory \cite{Ruppeiner1983a, Ruppeiner1983b, Diosi1985}. The $|R|\sim\xi^d$ proportionality has seen considerable computational success; see, for example, \cite{Ruppeiner1979, Johnston2003}. Note that $R$ is an intrinsically mesoscopic quantity, while interparticle potentials are microscopic. So the connection between $R$ and microscopic interaction potentials cannot be direct, and an attempt at a strict interpretation, without exceptions, could lead us astray. Perhaps the hard repulsive microscopic forces produce mesoscopic groups of atoms that resemble those produced by attractive microscopic forces.

\par
There is another point that must be kept in mind about the sign of $R$ in cases where $|R|$ is small. The first calculations of $R$ \cite{Ruppeiner1981, Ruppeiner1979} were done mostly near critical points, where $|R|^{1/3}$ is much larger than the average distance between atoms. Here, the volume measured by $|R|$ encompasses many atoms, and the dimensionless quantity $|R|\rho\gg 1$. One might expect that a meaningful physical interpretation of $R$ requires $|R|\rho\gg 1$. In \cite{Ruppeiner12a, Ruppeiner12b}, volumes with $|R|\rho\sim 1$ were styled as marking the low $|R|$ limit, and it was thought that $|R|$ much below this limit would be difficult to interpret physically. Nevertheless a number of calculations have been done below this limit, with findings that fit into a persistent and compelling picture. Nowadays the attitude is simply to calculate, and not to worry too much about the low $|R|$ limit.

\section{Conclusion \label{concl}}

The sign rule of the thermodynamic Ricci curvature scalar $R$ is perhaps the most important finding that arises from Riemannian thermodynamic geometry. It is expected that $R$ should be positive/negative for microscopic repulsive/attractive interactions. The role of repulsive and attractive interactions and their competition governs the form of the pair correlation function, i.e., the structure of the fluid and, subsequently, the macroscopic phase behavior. Therefore, exploring $R$ across the phase diagram of various fluids is significant since it allows the identification of phase space regions with predominantly repulsive or attractive behavior. 

\par
However, the interpretation of the $R$-sign rule is not as straightforward as previously thought and requires further in-depth analysis. In a recent study Bra\'nka, Pieprzyk, and Heyes \cite{Branka2018} found negative values for $R$ for the hard-sphere limit of the inverse power law fluid, where positive $R$ is generally expected. The reason for this anomalous behavior deviating from the $R$-sign rule is unclear. In this context, and because of the lack of a theoretical mesoscopic justification, further investigations of different potentials with a purely repulsive interaction might be interesting and could give additional information regarding the anomalous behavior. 

\par
In the present study we investigated the phase behavior of the GCM fluid by means of Riemannian geometry. In contrast to the "hard" inverse power law system discussed above, the GCM potential is an ultra-soft, bounded potential which is frequently used to describe colloidal systems. The main part of the phase diagram was analyzed with the aid of an EOS \cite{MayMau12b}, which was obtained from extensive computer simulations \cite{MauSad11} by fitting pressure data. In addition, the HTL and the HDL were examined with the help of the RPA, a theoretical approach that gives remarkably good results for the GCM in this region. A direct link to the BPH anomaly can be found in the LDL, hard-core regime. Here, the phase behavior of the GCM is compared with the Carnahan-Starling EOS \cite{Carnahan1969}, which is an excellent approximation for the hard-sphere gas.

\par
Our study shows that the thermodynamic Ricci curvature scalar $R$ is positive over most of the phase space, as is expected for a purely repulsive system. The bounded character of the GCM potential results in weaker structural correlations in the HTL and HDL, characterized by a pair-correlation function $g(r) \to 1$ . The system shows ideal gas-like behavior where $R = 0$ is expected. This behavior was impressively demonstrated in the context of the RPA. Furthermore, the freezing and melting line of the GCM fluid run into a common point at maximum freezing temperature $T_m$ with $\Delta v = 0$ between the solid and fluid phase. At this point, the possibility of a second-order critical point is frequently discussed in the literature \cite{Stillinger1976, Prestipino2005}. Thermodynamic metric geometry results in $R \to - \infty$ at a critical point. Since the $R$-values remain finite at this common point, a first-order phase transition must be present here in accordance with findings of ref. \cite{Stillinger1976, Prestipino2005}. Geometric analysis thus offers a tool to answer such questions.

\par
A comparably interesting region is the LDL. Although the GCM potential is purely repulsive, the analysis of the GCM-EOS showed negative $R$ in this region. Stillinger \cite{Stillinger1976, Stillinger1997} argued that the GCM acts here like a gas of hard-spheres, thus establishing a direct link to the BPH anomaly. We compared the behavior of the GCM with other hard sphere systems, e.g., on the basis of the Carnahan-Starling EOS. Additionally, within our discussion, we tried to give an explanation for this unusual behavior. Nevertheless, a fundamental theoretical interpretation of this phenomenon remains as an exciting issue for future research.

\section{Appendix}

In this Appendix, we explain the source of our GCM thermodynamics. Computer simulations \cite{MauSad11} produced pressure numbers that were fit to a functional form $p(T,\rho)$. This functional form, and the known ideal gas properties, yield the internal energy per volume $u(T,\rho)$ and the entropy per volume $s(T,\rho)$ \cite{MayMau12b}. The complete thermodynamics then follows from the Helmholtz free energy per volume $f(T,\rho)= u(T,\rho)-T s(T,\rho)$.

\par
In the derivation below, let us work with coordinates varying $(T,V)$, and holding $N$ fixed. The entropy $S=S(T,V)$ has differential,

\begin{equation} dS=\left(\frac{\partial S}{\partial T}\right)_V dT +\left(\frac{\partial S}{\partial V}\right)_T dV. \label{A10}\end {equation}

\noindent We have also the heat capacity at constant volume, 

\begin{equation} C_V = T\left(\frac{\partial S}{\partial T}\right)_V, \label{A20}\end{equation}

\noindent the first law of thermodynamics,

\begin{equation} dU=T\,dS-p\,dV, \label{A30}\end{equation}

\noindent and the Maxwell relation,

\begin{equation}\left(\frac{\partial S}{\partial V}\right)_T= \left(\frac{\partial p}{\partial T}\right)_{V}. \label{A40}\end{equation}

\noindent Eqs. (\ref{A10})-(\ref{A40}) yield the energy differential

\begin{equation}dU=C_V dT+\left[T\left(\frac{\partial p}{\partial T}\right)_{V}-p\right]dV. \label{A50}\end{equation}

\par
Writing $dV=-Nd\rho/\rho^2$, and integrating $dU$ along a curve of constant $T$ and $N$, starting from the ideal gas state $(\rho=0)$, to the state $(T,\rho)$, yields the internal energy per volume,

\begin{equation} u(T,\rho)=\displaystyle{\frac{3}{2}\,\rho T+\rho\int_{\rho=0}^{\rho}\left[p^{ex}-T\left(\frac{\partial p^{ex}}{\partial T}\right)_{\rho}\right]\frac{d\rho}{\rho^2}}. \label{A60}\end{equation}

\noindent We have assumed the ideal gas energy equation $U=\frac{3}{2} N T$, and separated the pressure into an ideal gas part $p^{id}=\rho\,T$, and a remaining excess part $p^{ex}$:

\begin{equation} p=p^{id}+p^{ex}.\label{A70}\end{equation}

\noindent Since $p^{id}$ cancels out in the integrand in Eq. (\ref{A60}), we have replaced $p$ by $p^{ex}$.

\par
The Sackur-Tetrode equation yields the ideal gas entropy $S=c N +\frac{3}{2}N\ln T - N\ln \rho$, where $c$ is a constant \cite{Landau1980}. Integrating the entropy differential Eq. (\ref{A10}) along a line of constant $N$ and $T$ now yields for the entropy per volume

\begin{equation}
s(T,\rho) = c \rho + \frac{3}{2}\rho\ln T - \rho\ln \rho - \rho\int^{\rho}_{\rho=0}\left( \frac{\partial p^{ex}}{\partial T} \right)_{\rho} \frac{d \rho}{\rho^{2}}.\label{A80}
\end{equation}

\noindent In the calculation of the thermodynamic Ricci curvature scalar $R$, the constant $c$ drops out, so it's value need not concern us here.

\par
If $T<T_{m}$, the integrations at constant $T$ in Eqs. (\ref{A60}) and (\ref{A80}) cross the solid region to reach a fluid state $B$ (see Fig. \ref{Figure1}) on the high-density side of the solid region. This suggests that to integrate correctly from $A\to B$ requires knowing either the thermodynamics in the solid region, or bypassing the solid region by integrating along some path such as $ACDB$ passing over the solid region. Both of these options present new challenges, but there is a simpler third option. Both $u(\rho,T)$ and $s(\rho,T)$ are expected to be analytic in the fluid state for all positive $\rho$ and $T$, because both the ideal gas equations of state and the fit formula for $p^{ex}(\rho,T)$ are analytic. Therefore, the straightforward integrations of $du$ and $ds$ over the curves $AB$ and $ACDB$ will produce the same values, despite the fact that the analytic extension of the fluid fit into the solid region does not model properly the physics in the solid region.

\par
As a consequence, the integrations in Eqs. (\ref{A60}) and (\ref{A80}) may be employed even if they run though the solid phase; namely, to get the thermodynamics at $B$ just integrate over the line $AB$. The same reasoning holds for the van der Waals gas where we must integrate though a coexistence region if $T$ is less than the critical temperature. The resulting equations of state are correct.

\section{DATA AVAILABILITY}

The data that support the findings of this study are available from the corresponding author upon reasonable request. This data includes the fit parameter set for determining the excess pressure.

\end{document}